\documentclass[twocolumn,english,aps,superscriptaddress,prb]{revtex4}
\pdfoutput=1
\usepackage[T1]{fontenc}
\usepackage[latin9]{inputenc}
\usepackage{color}
\usepackage{amsmath}
\usepackage{graphicx}
\usepackage{amssymb}
\usepackage{esint}
\usepackage{wasysym}
\usepackage[normalem]{ulem}

\makeatletter
\@ifundefined{textcolor}{}
{%
 \definecolor{BLACK}{gray}{0}
 \definecolor{WHITE}{gray}{1}
 \definecolor{RED}{rgb}{1,0,0}
 \definecolor{GREEN}{rgb}{0,1,0}
 \definecolor{BLUE}{rgb}{0,0,1}
 \definecolor{CYAN}{cmyk}{1,0,0,0}
 \definecolor{MAGENTA}{cmyk}{0,1,0,0}
 \definecolor{YELLOW}{cmyk}{0,0,1,0}
 }

\@ifundefined{definecolor}
 {\usepackage{color}}{}
\@ifundefined{definecolor}{\@ifundefined{definecolor}
 {\usepackage{color}}{}
}{}%\documentclass[aps,prb,preprint,showpacs]{revtex4}
\usepackage{epsf}

\newcommand{\yrs}{YbRh$_2$Si$_2$}
\newcommand{\ycs}{YbCo$_2$Si$_2$}
\newcommand{\yrsc}{Yb(Rh$_{0.73}$Co$_{0.27}$)$_2$Si$_2$}
\newcommand{\yrscx}{Yb(Rh$_{1-x}$Co$_{x}$)$_2$Si$_2$}
\newcommand{\ynp}{YbNi$_4$P$_2$}
\newcommand{\ynpax}{YbNi$_4$(P$_{1-x}$As$_x$)$_2$}

\newcommand{\TN}{T_{\rm N}}
\newcommand{\TC}{T_{\rm C}}
\newcommand{\TX}{T_{\rm X}}
\newcommand{\TL}{T_{\rm L}}

\makeatother

\usepackage{babel}

\begin{document}

%\title{Competing magnetic anisotropies and multicriticality: The case of Co-doped \yrs}
%\title{Magnetic frustration, competing anisotropies, and multicriticality:\\ The case of Co-doped \yrs}
\title{Competing orders, competing anisotropies, and multicriticality:\\ The case of Co-doped \yrs}

\author{Eric C. Andrade}
\affiliation{Institut f\"ur Theoretische Physik, Technische Universit\"at Dresden,
01062 Dresden, Germany}
\author{Manuel Brando}
\author{Christoph Geibel}
\affiliation{Max Planck Institute for Chemical Physics of Solids, N{\"o}thnitzer Str. 40, 01187~Dresden, Germany}
\author{Matthias Vojta}
\affiliation{Institut f\"ur Theoretische Physik, Technische Universit\"at Dresden,
01062 Dresden, Germany}

%%%%%%%%%%%%%%%%%%%%%%%%%%%%%%%%%%%%%%%%%%%%%%%%%%%%%%%%%%%%%%%%%%%%%%%

\begin{abstract}
Motivated by the unusual evolution of magnetic phases in stoichiometric and Co-doped
\yrs, we study Heisenberg models with
competing ferromagnetic and antiferromagnetic ordering combined with
competing anisotropies in exchange interactions and $g$ factors.
Utilizing large-scale classical Monte-Carlo simulations, we analyze the ingredients
required to obtain the characteristic crossing point of uniform susceptibilities observed
experimentally near the ferromagnetic ordering of \yrsc.
The models possess multicritical points, which we speculate to be relevant for the
behavior of clean as well as doped \yrs. We also make contact with experimental data on {\ynp}
where a similar susceptibility crossing has been observed.
%and furthermore discuss additional features arising from including a single-ion anisotropy.
\end{abstract}

\date{\today}


\maketitle
%%%%%%%%%%%%%%%%%%%%%%%%%%%%%%%%%%%%%%%%%%%%%%%%%%%%%%%%%%%%%%%%%%%%%%%

\section{Introduction}

Magnetic quantum phase transitions (QPT) in solids are an active area of research, with
many unsolved problems present in particular for metallic systems.\cite{geg08,hvl} While
the simplest theoretical models assume a single type of ordering instability and
SU(2) spin rotation symmetry, real materials often possess additional ingredients which
complicate the interpretation of experimental data:
(i) a competition of multiple ordered states, e.g. with different
ordering wavevectors and/or different spin structures, that may result from geometric
frustration or multiple nesting conditions, and
(ii) magnetic anisotropies that can modify ordered states and induce additional crossover
energy scales.

The tetragonal heavy-fermion metal {\yrs} is a prominent example for anisotropic and
competing magnetic orders. Its uniform magnetic susceptibility is by an order of
magnitude larger for fields $B_{ab}$ applied in the basal plane as compared to fields
$B_{c}$ along the c axis. Stoichiometric \yrs\ displays an ordered phase below
$\TN=70$\,mK, believed to be an antiferromagnet with moments oriented in the basal plane.
This order is destroyed at a field-driven QPT at $B_{ab}^{{\rm crit}}=60$\,mT, the
corresponding c axis critical field is $B_{c}^{{\rm
crit}}=0.66$\,T.\cite{trovarelli00,gegenwart02}

\begin{figure}[!t]
\includegraphics[width=0.35\textwidth]{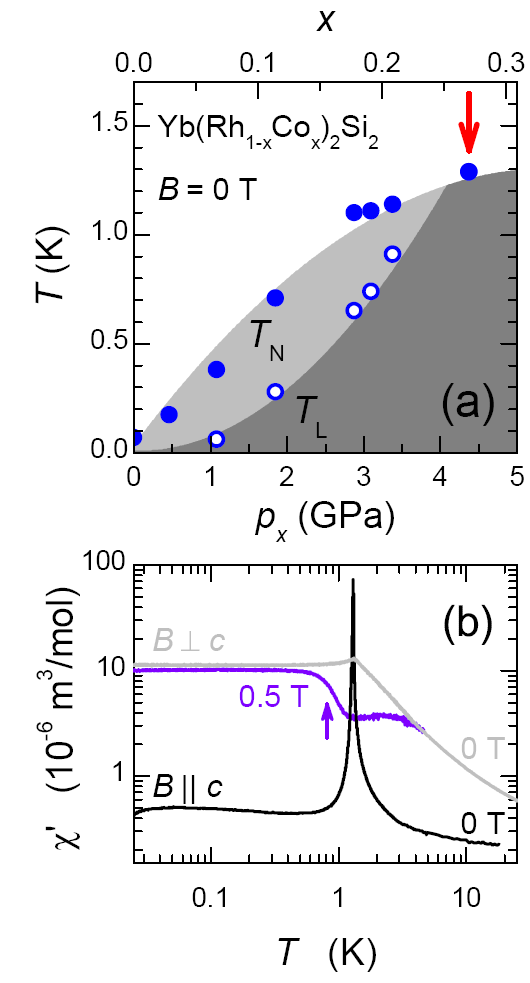}
\caption{ (a) Temperature-composition phase diagram of \yrs, showing the two magnetic
transitions at $\TN$ and $\TL$. Pressure (lower axis) has an effect similar to
Co doping (upper axis). (We note that ongoing studies suggest an even more complex phase diagram with
canted AFM order below $\TL$ for $x < 0.2$.)
(b) Susceptibility data from \yrsc, showing the switch of the anisotropy
of the magnetic response as function of temperature.
Figure reproduced from Ref.~\onlinecite{lausberg12}.
}
\label{fig:exp1}
\end{figure}

Upon applying hydrostatic pressure or doping with Co, both $\TN$ and $B^{{\rm crit}}$
increase which is primarily caused by a reduction of Kondo screening.
Moreover, a second phase transition at $\TL<\TN$ occurs at small fields,
Fig.~\ref{fig:exp1}(a), with the precise nature of the phase for $T<\TL$ being
unknown.\cite{friede09,klingner11}
Interestingly, an almost divergent uniform susceptibility was reported for stoichiometric
{\yrs}, hinting at the presence of ferromagnetic correlations.\cite{gegenwart05}
However, in-plane ferromagnetic order could be detected neither under pressure nor with
doping.

Given this state of affairs, it came as a surprise that \yrsc\ was recently
found\cite{lausberg12} to develop ferromagnetic order below $\TC=1.3$\,K with moments
oriented along the {\em c axis}, which was identified as the hard magnetic axis from all
previous measurements.
This ``switch'' of response anisotropies in \yrsc\ is
reflected in the susceptibilities, where $\chi_{ab}(T)$ and $\chi_{c}(T)$
display a characteristic crossing point at $\TX$ slightly above $\TC$,
Fig.~\ref{fig:exp1}(b). Furthermore, the evolution of $\TL$ with
doping in \yrscx, Fig.~\ref{fig:exp1}(a), suggests that the transition
at $\TL$ involves a ferromagnetic ordering component for $x<0.27$.
Parenthetically, we note that antiferromagnetism takes over again
for $x\gtrsim0.6$, with {\ycs} displaying a N\'{e}el transition at
$\TN=1.65$\,K and a large in-plane ordered moment of 1.4\,$\mu_{B}$.\cite{klingner11,klingner11b}

Taken together, the data show that the magnetism of {\yrscx} is determined by competing
antiferromagnetic (AFM) and ferromagnetic (FM) orders, which moreover are characterized by competing
anisotropies.

In this paper we propose that Heisenberg models of local moments can account for many of
the unusual magnetic properties of \yrscx, by invoking multiple magnetic interactions and
a competition between anisotropic exchange and anisotropic $g$ factors.
We primarily focus on the behavior of \yrsc\ in the vicinity of its finite-temperature
phase transition which can be modeled classically. For the simplest model, we
investigate thermodynamic properties using large-scale Monte Carlo (MC)
simulations, and provide an explicit comparison to experimental susceptibility data.
Based on our results, we propose that multicritical points, naturally present in our
modeling, are of relevance for understanding the unusual quantum criticality in
stoichiometric \yrs.
Finally, we also discuss the behavior of the heavy-fermion ferromagnet \ynp, which
displays a similar ``switch'' of magnetic response anisotropy.\cite{krellner11,steppke13}
%However, this material is located in the close vicinity of a quantum critical point, such
%that our classical modeling appears insufficient.

The body of the paper is organized as follows:
In Sec.~\ref{sec:model} we describe the family of anisotropic Heisenberg models to be
studied, together with methodological aspects of the MC simulations.
Sec.~\ref{sec:yrs} is devoted to a detailed modeling of \yrsc. It starts with a
discussion of competing magnetic anisotropies and then introduces competing FM
and AFM ordering tendencies, in order to match experimental observations.
In Sec.~\ref{sec:bicrit} we then discuss the presence of multicritical points in the phase
diagrams of our models and speculate on the role of quantum bicriticality in the phase
diagram of \yrs.
Finally, in Sec.~\ref{sec:ynp} we briefly discuss the case  of \ynp.
A short summary concludes the paper.
In the appendix we illustrate the effects of an additional single-ion anisotropy, not
present in the spin-1/2 models discussed in the bulk of the paper.

%%%%%%%%%%%%%%%%%%%%%%%%%%%%%%%%%%%%%%%%%%%%%%%%%%%%%%%%%%%%%%%%%%%%%%%

\section{Model and simulations}

\label{sec:model}

\subsection{Model and magnetic anisotropies}

The Yb$^{3+}$ ions in {\yrs} are in a $4f^{13}$ configuration
in a tetragonal crystal field. The ionic ground state is a $\Gamma^{7}$
Kramers doublet,\cite{sichelschmidt03} which can be represented by
an effective (pseudo)spin-1/2 moment, which further couples to conduction
electrons via a Kondo exchange coupling. Due to strong spin-orbital
coupling, SU(2) spin symmetry is broken, and magnetic anisotropies
appear: (i) The (pseudo)spins couple to an external Zeeman field in
an anisotropic fashion, leading to an anisotropic $g$ tensor. (ii)
Magnetic interactions between the moments will be anisotropic in spin
space -- this applies to both direct exchange and RKKY interactions.
A single-ion anisotropy is forbidden for elementary spin-1/2; for
completeness we discuss its effects in models of spins $S\geq1$ in
the appendix.

In this paper, we consider spin-only models of Yb moments in \yrs.
These models will be of Heisenberg type; to account for competing
phases, we will include longer-range exchange interactions. For simplicity,
we place the moments on a cubic lattice, and consider models of the
form
\begin{equation}
\mathcal{H}=-\sum_{\left\langle i,j\right\rangle
\alpha}J_{1}^{\alpha}S_{i}^{\alpha}S_{j}^{\alpha}-\sum_{\left\langle \langle
i,j\right\rangle
\rangle\alpha}J_{2}^{\alpha}S_{i}^{\alpha}S_{j}^{\alpha}-\sum_{i\alpha}g_{\alpha}h_{\alpha}S_{i}^{\alpha}
\label{h}
\end{equation}
where $J_{1}$ and $J_{2}$ denote couplings to first and second
neighbors, and $S_{i}^{\alpha}$ is the $\alpha=x,y,z$ component
of the moment at site $i$, $h_{\alpha}$ represents the external
magnetic field, and $g_{\alpha}$ are the (diagonal) components of
the $g$ tensor; in a tetragonal environment, we have $g_{x}=g_{y}\equiv g_{ab}$
and $g_{z}\equiv g_{c}$. The anisotropic exchange interaction can
be parameterized as $J_{1}^{x}=J_{1}^{y}=J_{1}\left(1+\Delta_{1}\right)$
and $J_{1}^{z}\equiv J_{1}$. For simplicity, we consider isotropic
second-neighbor couplings, with $J_{2}^{\alpha}=J_{2}$. Being interested
in the behavior near the finite-temperature phase transition, we approximate
the quantum spins as classical vectors with $|\vec{S_{i}}|=1$.

%%%%%%%%%%%%%%%%%%%%%%%%%%%%%%%%%%%%%%%%%%%%%%%%%%%%%%%%%%%%%%%%%%%%%%%

\subsection{Monte-Carlo simulations}

To access the finite-temperature behavior of \eqref{h} we perform
equilibrium MC simulations on cubic lattices of size
$N=L\times L\times L$, with $L\leq32$ and periodic boundary conditions.
We employ the single-site Metropolis algorithm combined with microcanonical
steps to improve the sampling at lower temperatures. Typically, we
run $10^{6}$ MC steps (MCS) as initial thermalization followed
by $10^{6}$ MCS to obtain thermal averages, which are calculated
by dividing the measurement steps into $10$ bins. Here one MCS corresponds
to one attempted spin flip per site. We consider the same number of
Metropolis and microcanonical steps. In all
our results we set $k_{B}=1$ and $a=1$, where $k_{B}$ is the Boltzmann
constant and $a$ is the lattice spacing and use $J_{1}$ as our energy
scale.

From the MC data, we calculate thermodynamic observables, as the specific
heat $C$ and the uniform magnetic susceptibilities, both along the
c axis
\begin{eqnarray}
\chi_{c}\left(T\right) & = & g_{c}^{2}\frac{N}{T}\left(\left\langle m_{c}^{2}\right\rangle -\left\langle m_{c}\right\rangle ^{2}\right),
\label{eq:chi_c}
\end{eqnarray}
and in the ab plane
\begin{eqnarray}
\chi_{ab}\left(T\right) & = & \frac{1}{2}g_{ab}^{2}\frac{N}{T}\left\langle
m_{ab}^{2}\right\rangle ,
\label{eq:chi_ab}
\end{eqnarray}
where $\left\langle \cdots\right\rangle $ denotes MC average, $m_{c}=M_{z}$,
and $m_{ab}=(M_{x}^{2}+M_{y}^{2})^{1/2}$ with $M_{\alpha}=N^{-1}\sum_{i}S_{i}^{\alpha}$.
Due to its definition, $m_{ab}$ incorporates the fact that the $x$
and $y$ components of $\vec{S}$ are equivalent.
In $\chi_{ab}$ no subtraction of expectation values is necessary, as we will only
consider ferromagnetic states with order along the $c$ axis. In the linear-response limit
$g$ factors do not enter the MC simulations directly, but appear as prefactors
in Eqs.~\eqref{eq:chi_c} and \eqref{eq:chi_ab} only.

Further we analyze phase transitions via the Binder cumulant,\cite{landau_binder}
and we characterize ordered states by the static spin structure factors
\begin{eqnarray}
S_{c}\left(\vec{q}\right) & = & \frac{1}{N}\sum_{i,j}e^{-i\vec{q}\cdot\vec{r}_{ij}}\left\langle S_{i}^{z}S_{j}^{z}\right\rangle ,\label{sfaczz}\end{eqnarray}
 \begin{eqnarray}
S_{ab}\left(\vec{q}\right) & = & \frac{1}{N}\sum_{i,j}e^{-i\vec{q}\cdot\vec{r}_{ij}}\left\langle S_{i}^{x}S_{j}^{x}+S_{i}^{y}S_{j}^{y}\right\rangle .\label{sfacperp_MC}\end{eqnarray}
 In a long-range ordered phase: $S_{c\left(ab\right)}\left(\vec{q}\right)/N\rightarrow m_{c\left(ab\right)}^{2}\delta_{\vec{q},\vec{Q}}$,
where $\delta$ is the Kronecker delta and $\vec{Q}$ is the ordering
wavevector. Note that $m_{c\left(ab\right)}\to1$ as $T\to0$, because
the classical ground state is fluctuationless. The behavior of $S_{c\left(ab\right)}(\vec{q})$
near $\vec{Q}$ allows to extract the correlation length $\xi_{c\left(ab\right)}\left(T\right)$
characterizing the magnetic order. In our finite-size simulations,
the ordering temperature for the second-order phase transition $\TC$
is efficiently extracted from the crossing points of $\xi(T)/L$
data for different $L$, according to the scaling law $\xi\left(T\right)/L=f(L^{1/\nu}(T-\TC))$,
where $f(x)$ is a scaling function and $\nu$ is the correlation
length exponent. % Since both \eqref{sfaczz} and \eqref{sfacperp_MC}
%are disconnected correlation function, the correlation length obtained
%in this fashion is only related to the physical correlation length
%at $T>\TC$.

%%%%%%%%%%%%%%%%%%%%%%%%%%%%%%%%%%%%%%%%%%%%%%%%%%%%%%%%%%%%%%%%%%%%%%%
%%%%%%%%%%%%%%%%%%%%%%%%%%%%%%%%%%%%%%%%%%%%%%%%%%%%%%%%%%%%%%%%%%%%%%%
%%%%%%%%%%%%%%%%%%%%%%%%%%%%%%%%%%%%%%%%%%%%%%%%%%%%%%%%%%%%%%%%%%%%%%%

\section{Modeling of Y\lowercase{b}(R\lowercase{h}$_{0.73}$C\lowercase{o}$_{0.27}$)$_{2}$S\lowercase{i}$_{2}$}

\label{sec:yrs}

In this section, we describe a stepwise modeling of magnetism in {\yrsc}
using spin models of the form \eqref{h}, with the goal of reproducing
the temperature dependence of $\chi(T)$ in Fig.~\ref{fig:exp1}(b).
Conduction electrons are not part of this modeling, i.e., we assume that the magnetic
properties can be described in terms of (effective) local-moment models. We note that
this assumption is not in contradiction with possible Kondo screening of the Yb 4f
moments: First, in Kondo-lattice systems the local-moment contributions to the magnetic
response are typically much larger than those of the conduction electrons. Second,
universality dictates that, even in cases where magnetism is fully itinerant, local-moment
models successfully describe many long-wavelength magnetic properties.\cite{mnsi_descr}

\subsection{Competing anisotropies and bicriticality}

\label{sec:fonly}

A central observation in the phase diagram\cite{klingner11,lausberg12}
of \yrscx, Fig. \ref{fig:exp1}a, is that the two thermal phase transitions
at $\TN$ and $\TL$, existing for small $x$, merge close to $x=0.27$.
Combining the facts that $\TN$ marks a transition into an in-plane
antiferromagnet and the state below $\TC$ at $x=0.27$ is a c-axis
ferromagnet suggests the competition of these two instabilities, with
a proximate bicritical (or, more generally, multicritical) point.

\begin{figure*}[t]
\includegraphics[width=0.98\textwidth]{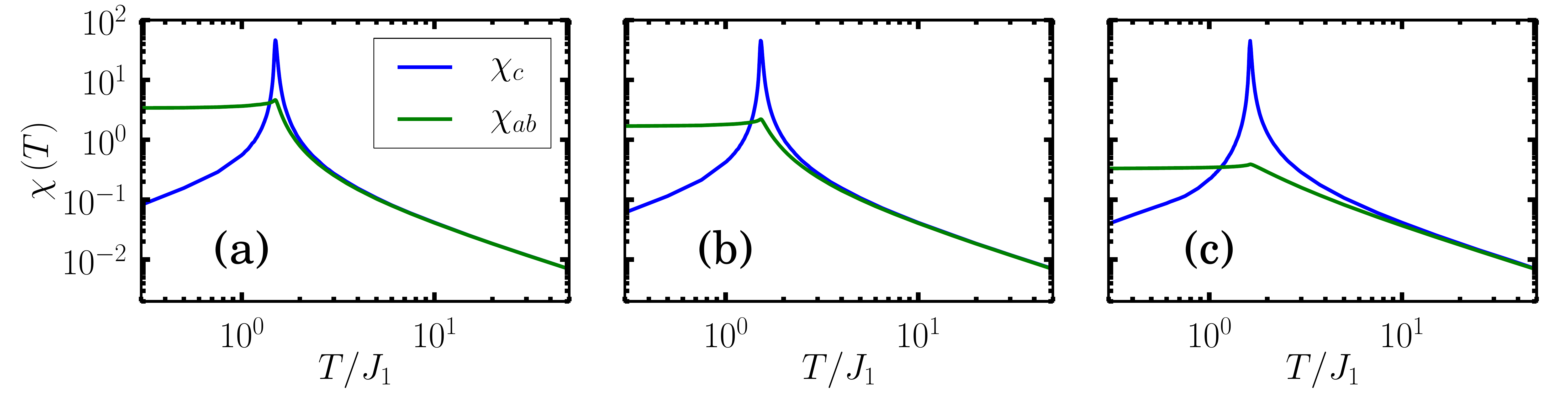}
\caption{\label{fig:chi-raw}
Anisotropic susceptibilities $\chi_{ab}$ (green) and $\chi_{c}$ (blue) on a
log-log scale for FM ($J^{x,y}>0$) spin exchange in the ab plane, $J_{2}=0$,
$g_{ab}=g_{c}$, and $L=16$. (a) $\Delta=-0.05$; (b) $\Delta=-0.10$; (c) $\Delta=-0.50$.
Finite-size effects are negligible except for temperatures within 0.2\% of $\TC$.
}
\end{figure*}

We start by focusing on the competition between in-plane and c-axis
order, which can be captured, in a spin-1/2 model as in Eq.~\eqref{h}, by anisotropic
exchange couplings, with $|J_{1}^{x,y}|=|J_{1}^{z}|(1+\Delta_{1})$.
Degeneracy of the two orders is trivially reached at the Heisenberg
point, $\Delta_{1}=0$, where the transition corresponds to
a bicritical point. A transition to c-axis order as in {\yrsc}
requires $\Delta_1<0$. Sample susceptibilities
results assuming FM ($J^{x,y}>0$) spin exchange in the ab plane are
shown in Fig.~\ref{fig:chi-raw}, where we see that while $\chi_{c}\left(T\right)$
is essentially independent of the anisotropy $\Delta_{1}$, $\chi_{ab}\left(T\right)$
is considerably enhanced as we approach the bicritical point, due
to the increase in the in-plane fluctuations. The behavior of the
magnetic susceptibilities in the vicinity of a bicritical point is
further explored in Appendix \ref{sec:appendix A}.

The fact that, experimentally, $\chi_{c}\ll\chi_{ab}$ for a large
range of temperatures above the transition (essentially from $\TC$ up to 100\,K)
implies $g_{c}<g_{ab}$, i.e., the g-factor anisotropy is opposite
to the exchange anisotropy. Indeed, for undoped \yrs\ a strong g-factor
anisotropy of $g_{ab}/g_{c}\approx20$ has been deduced from electron-spin-resonance
(ESR) experiments.\cite{sichelschmidt03} For {\yrsc} we have\cite{lausberg12} $g_{ab}/g_{c}=6.4$
and we note that the end member of the doping series, $\mbox{Yb}\mbox{Co}_{2}\mbox{Si}_{2}$,
has a smaller anisotropy, $g_{ab}/g_{c}\approx2.5$, indicating a
considerable variation of the g-factor anisotropy ratio throughout
the series, even though no change in the symmetry of the ground-state
doublet is observed.\cite{gruner12,klingner11b}

In a modeling of the susceptibility data of {\yrsc} under the assumption of exclusively
nearest-neighbor ferromagnetic coupling $J_{1}^{\alpha}>0$ (this assumption will be
relaxed in Sec.~\ref{sec:faf} below) we are left with the following free parameters: the
overall energy scale $J_{1}$ which determines $\TC$, the exchange anisotropy $\Delta$,
and the g-factor anisotropy $g_{ab}/g_{c}$. (The absolute value of $\chi$, or $g$, is simply
adjusted to the data by matching the low-$T$ value of $\chi_{ab}$.)
To determine $\Delta$ and $g_{ab}/g_{c}$ different strategies appear possible; we found
the following useful: First, $\Delta$ is chosen to match the temperature dependence of
the non-divergent $\chi_{ab}(T)$ near $\TC$ (where $\chi_{c}$ diverges). In general,
small $\Delta$ should to be chosen to ensure proximity to bicriticality. Second, with
$\Delta$ fixed, the crossing temperature $\TX$ of $\chi_{ab}$ and $\chi_{c}$ can be used
to determine the $g$ factor anisotropy by demanding that experiment and theory yield the
same $\TX/\TC$ ratio; experimentally this number is $\TX/\TC=1.016$.
After this procedure is implemented, we add a constant susceptibility $\chi_{v}$ to mimic
the van-Vleck contribution arising from higher crystal-electric-field levels which are
not part of our model. The value chosen is $\chi_{v}=0.20\times10^{-6}\mbox{
m}^{3}\mbox{/mol}$, which is close to the value of $0.172\times10^{-6}\mbox{
m}^{3}\mbox{/mol}$ deduced for \yrsc.\cite{lausberg12}

\begin{figure}[!t]
\includegraphics[width=0.47\textwidth]{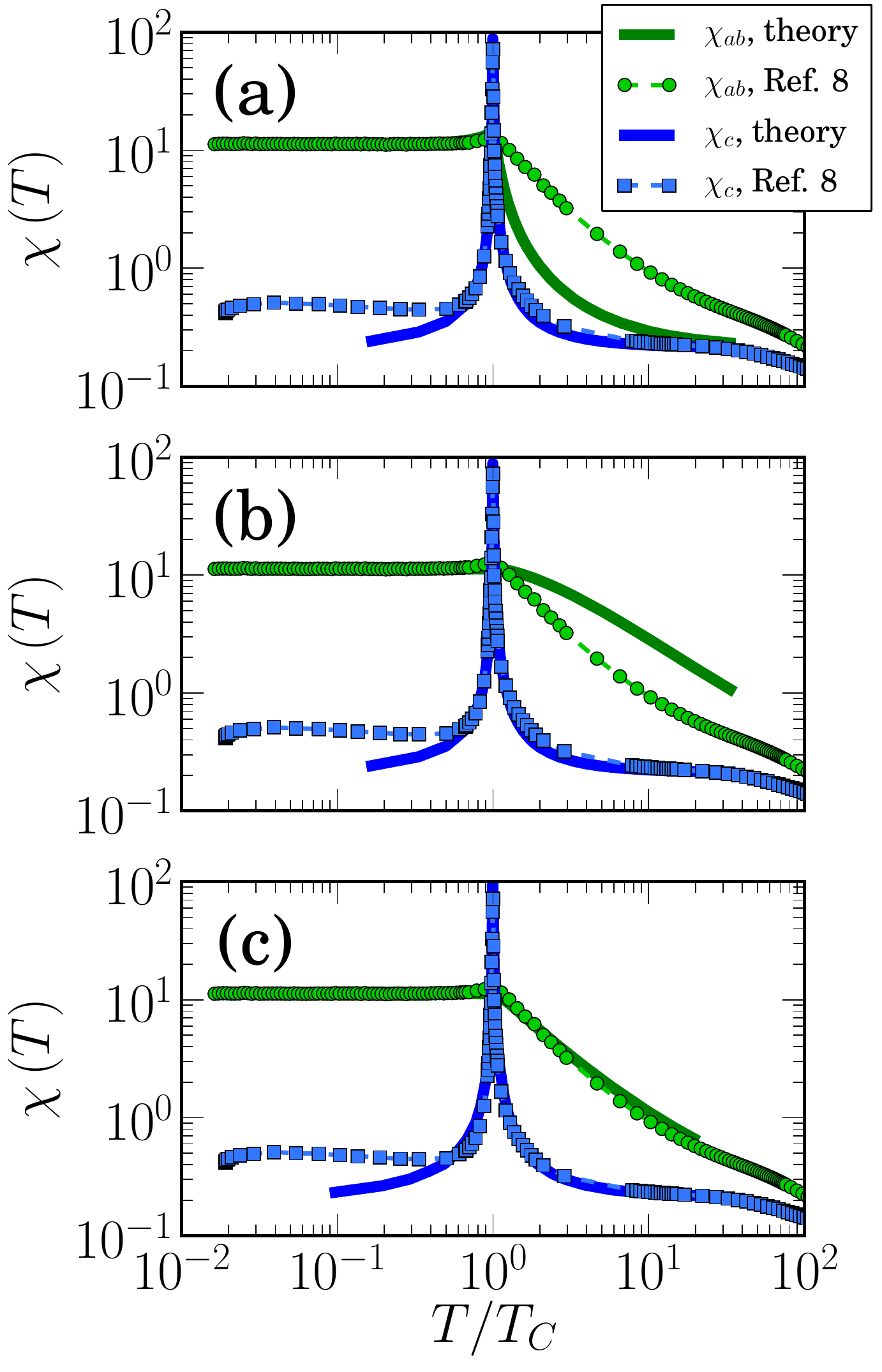}
\caption{\label{fig:comp-ganiso}
Comparison between the numerical and experimental susceptibilities $\chi(T)$ as a
function of $T/\TC$ on a log-log scale. The continuous curves show the
simulation results, $\chi_{ab}$ in green and $\chi_c$ in blue, obtained for
$\Delta=-0.05$ and $L=32$ in Eq. \eqref{h}, with different exchange parameters in panels
(a)-(c). The dashed lines with symbols represent the corresponding experimental
susceptibilities reported in Ref.~\onlinecite{lausberg12} in units of $10^{-6}\mbox{
m}^{3}\mbox{/mol}$, with $\TC=1.30\mbox{ K}$.
(a) FM ($J^{x,y}>0$) spin exchange in the ab plane, with $g_{ab}/g_{c}=2.5$ and $J_{1}=0.87\mbox{ K}$.
(b) Same as (a), but for AFM ($J^{x,y}<0$) spin exchange in the ab plane and $g_{ab}/g_{c}=16.4$.
(c) Same as (b), with additional isotropic next-nearest neighbor FM coupling $J_{2}$, with
$J_{2}=J_{1}$, $J_{1}=0.25\mbox{ K}$ and $g_{ab}/g_{c}=7.3$.
In all simulation results, we have added a constant van-Vleck background
$\chi_{v}=0.20\times10^{-6}\mbox{ m}^{3}\mbox{/mol}$, as is seen in
\yrsc.\cite{lausberg12}
}
\end{figure}

For exclusively nearest-neighbor ferromagnetic coupling, the first step cannot be
satisfactorily followed: The temperature variation of $\chi_{ab}$ is too large compared
to experiment. A sample set of data is presented in Fig. \ref{fig:comp-ganiso}a. Here
$\Delta=-0.05$ has been chosen as a compromise; fixing $\TX/\TC$ then yields
$g_{ab}/g_{c}=2.5$. The poor agreement with experiment will be cured upon considering
dominant in-plane {\em antiferromagnetic} correlations below.

%%%%%%%%%%%%%%%%%%%%%%%%%%%%%%%%%%%%%%%%%%%%%%%%%%%%%%%%%%%%%%%%%%%%%%%

\subsection{Competing ferromagnetism and antiferromagnetism}

\label{sec:faf}

While the significant increase of $\chi_{ab}$ towards low temperatures
in both Co-doped and undoped {\yrs} reflects the presence of ferromagnetic
exchange, the tendency towards antiferromagnetism cannot be ignored.
Since undoped {\yrs} is believed (and {\ycs} was proved) to display low-temperature
antiferromagnetic order, it is very likely that strong in-plane antiferromagnetic
exist in {\yrsc} as well.

First we repeat the analysis described in Sec.~\ref{sec:fonly} with purely AFM in-plane
exchange, i.e., we consider the model \eqref{h} with $J_{1}^{x,y}=-J_{1}(1+\Delta_{1})<0$
and $J_{1}^{z}=J_{1}>0$. Corresponding results are presented in
Fig.~\ref{fig:comp-ganiso}b. Since the coupling of the spins' $z$ components remains FM,
$\chi_{c}$ does not change as compared to Fig.~\ref{fig:comp-ganiso}a. In contrast,
$\chi_{ab}$ is not only strongly suppressed but also shows much less temperature
variation, as expected for an antiferromagnet: For $T>\TC$ the susceptibility follows
roughly $\chi_{ab} \propto 1/(T-\Theta_{ab})$ with $\Theta_{ab}\sim J^{x,y}<0$, which results in a
flattening towards low temperature.
To restore the $\chi$ crossing at the experimental value of $\TX/\TC$ we now need to
assume a $g$-factor ratio of $g_{ab}/g_{c}=16.4$. The behavior of $\chi_{ab}$ for $T<\TC$
is now much closer to the one observed in \yrsc. However, its very slow decay upon
heating above $\TC$ still indicates a rather poor match with the experiment.

The natural conclusion is that both FM and AFM tendencies are present
for the in-plane order, a qualitative conclusion reached earlier on
the basis of experimental data.\cite{gegenwart05} There are various
ways to implement the competition of FM and AFM order in a microscopic
spin model. Here we choose a simple and frustration-free way: We employ
an isotropic ferromagnetic second-neighbor coupling $J_{2}$, such
that at $J_{1}^{xy}=0$ the in-plane ordered states with wavevectors
${\vec{Q}}=(0,0,0)$ and ${\vec{Q}}=(\pi,\pi,\pi)$ are degenerate,\cite{collinear}
and $J_{1}^{xy}\gtrless0$ can then be used to tip the balance between
dominant ferromagnetic and antiferromagnetic in-plane correlations.

Results with $J_{2}$ included are presented in Fig. \ref{fig:comp-ganiso}c.
As anticipated, a $J_{2}^{xy}>0$ enhances the FM fluctuations in
the ab plane, which has two immediate effects: (i) The $g_{ab}/g_{c}$
ratio required to fit $\TX/\TC$ decreases with increasing $J_{2}$
and (ii) the behavior of $\chi_{ab}$ for $T>\TC$ also approaches
the one from $\chi_{ab}^{exp}$ as we increasing $J_{2}$, indicating
that the effects of this extra exchange term are considerably felt
for $T\gtrsim\TC$.

Whereas for FM in-plane correlation $\chi_{ab}\left(T\right)$ shows
a distinct dependence on the anisotropy $\Delta_{1}$, Fig. \ref{fig:chi-raw},
this effect is considerably weaker for AFM correlations since in this
case $\chi_{ab}$ is itself strongly suppressed. Therefore, we obtain
similar results, as in Fig. \ref{fig:comp-ganiso}c, for $\Delta_{1}=-0.10$
and even $\Delta_{1}=-0.50$. While this diminishes our predictive power
w.r.t. the precise value of the anisotropy, we recall that Fig.~\ref{fig:exp1}b
suggests the proximity to bicritical point, which
constrains $\Delta_{1}$ to be small.

The good match of theoretical and experimental susceptibilities in
Fig. \ref{fig:comp-ganiso}c suggests that our classical anisotropic
Heisenberg model captures the essentials of the ferromagnetic ordering
process in \yrsc.

We note that the scenario of different competing kinds of in-plane exchange interactions
is also (indirectly) supported by the complicated propagation vectors observed in pure {\ycs}, $\vec{Q} =
2\pi (1/4, 0.08, 1)$ and $\vec{Q} = 2\pi (1/4, 1/4, 1)$ for the high-$T$ and low-$T$
phases, respectively. While such $\vec{Q}$ might also result from competition of purely
AFM exchanges, our analysis of the $T$ dependence of $\chi_{ab}$ for $T > \TC$ gives a
very strong hint that some of these in-plane exchanges have to be ferromagnetic.

%%%%%%%%%%%%%%%%%%%%%%%%%%%%%%%%%%%%%%%%%%%%%%%%%%%%%%%%%%%%%%%%%%%%%%%
%%%%%%%%%%%%%%%%%%%%%%%%%%%%%%%%%%%%%%%%%%%%%%%%%%%%%%%%%%%%%%%%%%%%%%%
%%%%%%%%%%%%%%%%%%%%%%%%%%%%%%%%%%%%%%%%%%%%%%%%%%%%%%%%%%%%%%%%%%%%%%%

\section{Multicriticality: Classical and quantum}
\label{sec:bicrit}

The modeling of {\yrsc} presented so far suggests the relevance of bicritical behavior,
arising from the competition of in-plane antiferromagnetism and out-of-plane
ferromagnetism. Indeed, the temperature--doping phase diagram of \yrs,
Fig.~\ref{fig:exp1}, indicates the presence of a finite-temperature bicritical point at a
doping level of $x\approx 25$\%. In the simplest scenario, the lower-temperature phase
transition line emerging from this point, labeled $\TL$, would then correspond to a
first-order transition between antiferromagnetic and ferromagnetic phases, but a more
complicated structures with mixed (e.g. canted) orders are possible as well (those are
not described by the simple Heisenberg models of Sec.~\ref{sec:model}).

A remarkable feature of the $\TN$ and $\TL$ lines in Fig.~\ref{fig:exp1} is that $\TL$,
although more strongly suppressed with decreasing $x$ as compared to $\TN$, appears to be
finite almost down to $x=0$. Given that $\TN$ is almost vanishing there as well, this
invites speculations about (approximate) {\em quantum bicriticality} near $x=0$.
This scenario would involve simultaneous quantum criticality of both itinerant in-plane
antiferromagnetism and c-axis ferromagnetism. In fact, some observations in
stoichiometric (or slightly Ge-doped) \yrs\ have been suggested to be compatible with
{\em ferromagnetic} quantum criticality: This includes the $T^{-0.7}$ dependence of the
Grueneisen parameter $\Gamma$ and the $T^{-0.3}$ dependence of the specific heat $C/T$ at
low temperatures above $\TN$ (see Ref.~\onlinecite{hvl}). However, this agreement has
mainly been considered fortuitous, as the ordered phase of stoichiometric \yrs\ is not
ferromagnetic. Quantum bicriticality would thus provide a new angle on the observations.

On the theoretical side, we note that bicriticality involving in-plane and
c-axis order by itself is not exotic, but simply corresponds to a point with an emergent
higher symmetry, here O(3).
What is, however, interesting and potentially exotic is quantum bicriticality involving
{\em itinerant} antiferromagnetism and ferromagnetism, which has not been studied
theoretically. Within the Landau-Ginzburg-Wilson description, pioneered by
Hertz,\cite{hertz76} these two transitions have different dynamical exponents, $z=2$
and 3, respectively, such that one can expect rich and non-trivial crossover phenomena
even if both transitions are above their respective upper-critical
dimension.\cite{meng12} This defines a fascinating field for future theoretical studies.\cite{dimxnote}

We note that another possible candidate for such a quantum bicriticality scenario in an itinerant system is the Laves phase NbFe$_{2}$, where competing AFM and FM orders have been observed near the quantum critical point.\cite{moroni2009}

%%%%%%%%%%%%%%%%%%%%%%%%%%%%%%%%%%%%%%%%%%%%%%%%%%%%%%%%%%%%%%%%%%%%%%%
%%%%%%%%%%%%%%%%%%%%%%%%%%%%%%%%%%%%%%%%%%%%%%%%%%%%%%%%%%%%%%%%%%%%%%%
%%%%%%%%%%%%%%%%%%%%%%%%%%%%%%%%%%%%%%%%%%%%%%%%%%%%%%%%%%%%%%%%%%%%%%%

\section{Comparison to Y\lowercase{b}N\lowercase{i}$_4$P$_2$}
\label{sec:ynp}

Our success in modeling the finite-temperature susceptibilities of {\yrsc} suggests to
look at other materials with similar phenomenology. Remarkably, the heavy-fermion
ferromagnet {\ynp} has been recently found\cite{krellner11,steppke13} to display a switch of
magnetic response anisotropy similar to \yrsc.

{\ynp} is a tetragonal metal with a Curie temperature of $\TC = 0.15$\,K. The
low-temperature ferromagnetic moment points perpendicular to the c axis, as indicated by
the divergent susceptibility $\chi_{ab}$ at $\TC$. Notably, $\chi_c(T)$ is larger than
$\chi_{ab}(T)$ for essentially all temperatures above $\TC$. Hence, the $g$ factors obey
$g_c > g_{ab}$, such that both the exchange and $g$ factor anisotropies appear opposite
to those of \yrsc. In \ynp, these anisotropies lead to a crossing of $\chi_{ab}(T)$ and
$\chi_c(T)$ at $\TX/\TC \approx$ 1.13.

There is, however, an additional ingredient relevant for {\ynp} (in contrast to \yrsc):
This material appears to be located extremely close to a quantum critical point: $\TC$
can be suppressed with a small amount of As doping in {\ynpax}, with $\TC\to0$
extrapolated for $x\approx 0.1$. This nearly {\em quantum} critical behavior in {\ynp} is
manifest in the temperature dependence of $\chi$, which follows $\chi_c\propto T^{-0.66}$
in the temperature range $\TC<T<$\,10\,K.
It is this power law which cannot be reproduced by any means is our classical simulation:
As seen in Fig.~\ref{fig:chi-raw} the high-temperature behavior of $\chi$ is Curie-like,
and this divergence becomes stronger upon approaching $\TC$. In contrast, in {\ynp} the
high-temperature Curie-like behavior, $\chi_c\propto 1/T$, is followed by the weaker
power-law divergence upon cooling.

Recalling the good match we were able to find for {\yrsc} we conclude that the latter
material is dominated by effectively classical magnetism, and the quantum critical
behavior of stoichiometric {\yrs} does not appear to extend up to $27\%$ of Co doping.

%%%%%%%%%%%%%%%%%%%%%%%%%%%%%%%%%%%%%%%%%%%%%%%%%%%%%%%%%%%%%%%%%%%%%%%
%%%%%%%%%%%%%%%%%%%%%%%%%%%%%%%%%%%%%%%%%%%%%%%%%%%%%%%%%%%%%%%%%%%%%%%
%%%%%%%%%%%%%%%%%%%%%%%%%%%%%%%%%%%%%%%%%%%%%%%%%%%%%%%%%%%%%%%%%%%%%%%

\section{Summary}
\label{sec:concl}

Motivated by the doping and temperature evolution of magnetism of \yrscx, we studied
Heisenberg models with competing anisotropies in the exchange interactions and $g$ factors.
Such models possess ordered phases with spins aligned either along the c axis or in the
ab plane, separated from each at low temperature by a first order line which ends at a
bicritical point. The proximity to this bicritical point, combined with the competition
between ferromagnetic and antiferromagnetic ordering, naturally explains the ``switch''
in the anisotropy in the magnetic susceptibilities as experimentally observed in \yrsc.

While our classical local-moment modeling gives a good account on the susceptibility data
of {\yrsc} near its finite-temperature phase transition, it fails to describe {\ynp}
where a similar switch of response anisotropies has been observed. This highlights the
importance of quantum fluctuations, as {\ynp} is located very close to a putative
ferromagnetic quantum critical point.

More broadly, our results point towards bicritical behavior being relevant in certain
Yb-based heavy-fermion compounds. Metallic quantum bicritical points, in particular
involving phenomena with different critical exponents, are thus an exciting field for
future research.

%%%%%%%%%%%%%%%%%%%%%%%%%%%%%%%%%%%%%%%%%%%%%%%%%%%%%%%%%%%%%%%%%%%%%%%

\acknowledgments

We are grateful to S. Lausberg for providing experimental data and for discussions.
ECA moreover thanks F. A. Garcia for discussions.
This research was supported by the DFG through FOR 960 and GRK 1621.
We also acknowledge ZIH at TU Dresden for the allocation of computing resources.

%%%%%%%%%%%%%%%%%%%%%%%%%%%%%%%%%%%%%%%%%%%%%%%%%%%%%%%%%%%%%%%%%%%%%%%
%%%%%%%%%%%%%%%%%%%%%%%%%%%%%%%%%%%%%%%%%%%%%%%%%%%%%%%%%%%%%%%%%%%%%%%
%%%%%%%%%%%%%%%%%%%%%%%%%%%%%%%%%%%%%%%%%%%%%%%%%%%%%%%%%%%%%%%%%%%%%%%

\begin{figure}[t]
\includegraphics[width=0.45\textwidth]{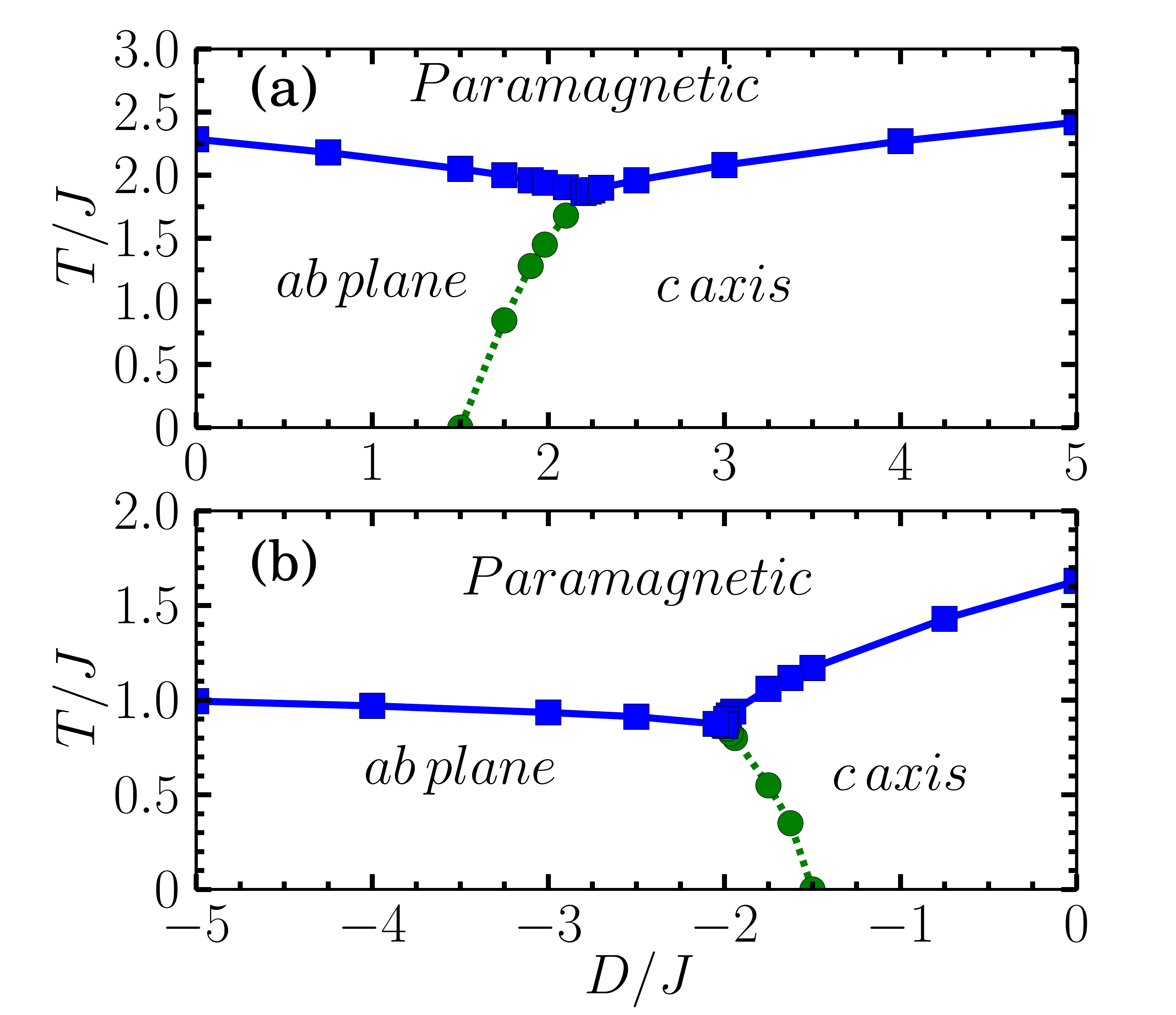}
\caption{\label{fig:Phase-diagram}
Phase diagram of model \eqref{h2} for fixed
values of the exchange anisotropy: $\Delta=0.5$ in (a) and $\Delta=-0.5$
in (b). The squares represent second-order phase transitions and the
circles first order phase transitions between the two ordered states.
The first order line ends at a bicritical point located at $D=2.220(2)$
(a) and $D=-1.995(5)$ (b).
}
\end{figure}

\begin{figure*}[t]
\includegraphics[width=0.9\textwidth]{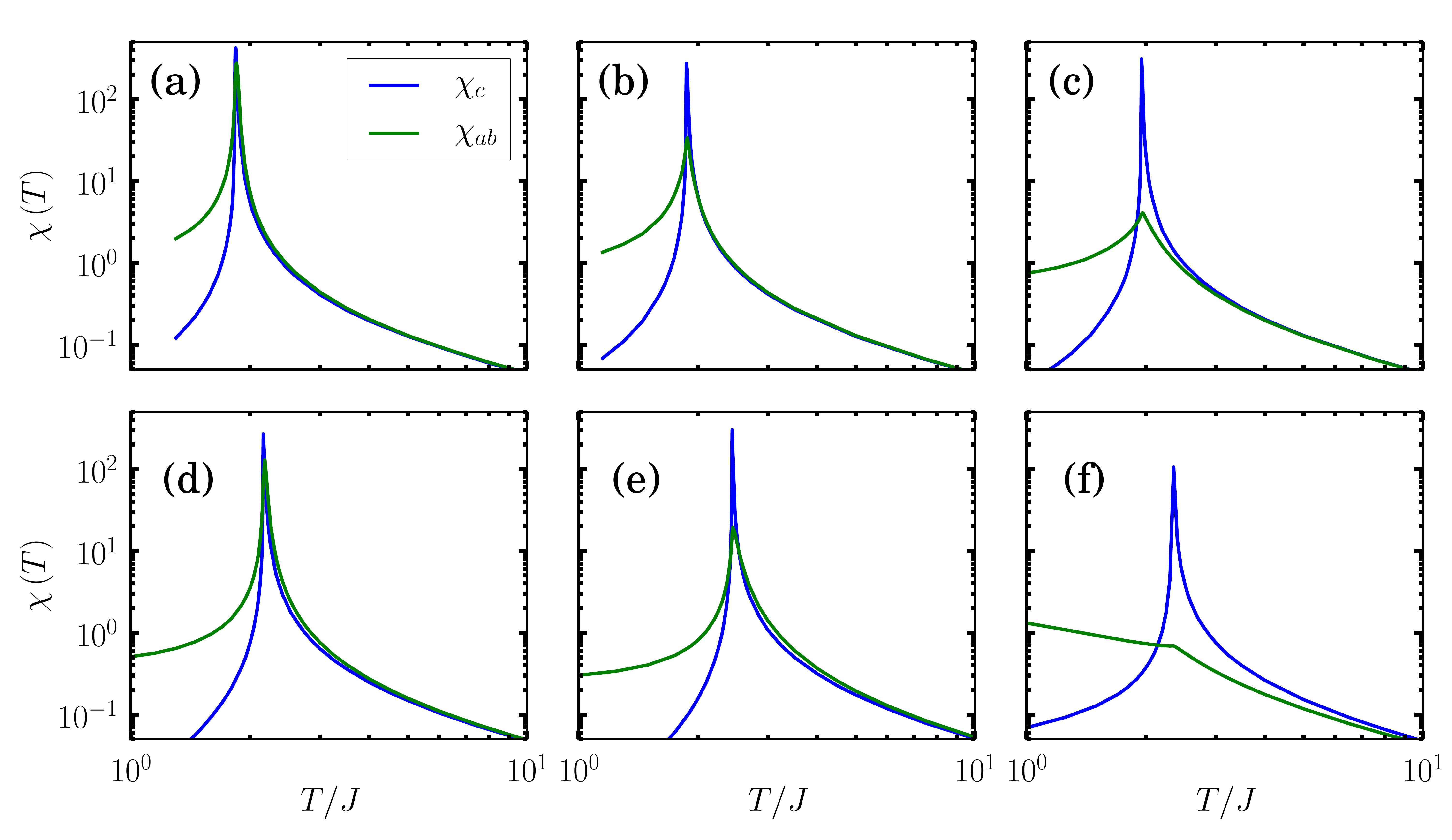}
\caption{\label{fig:chi}
Anisotropic susceptibilities $\chi_{ab}$ and $\chi_{c}$
on a log-log scale for $g_{ab}=g_{c}$ and $L=32$. (a) $\Delta=0.5$
and $D=2.20J$; (b) $\Delta=0.5$ and $D=2.25J$; (c) $\Delta=0.5$
and $D=2.50J$; (d) $\Delta=1.0$ and $D=4.40J$; (e) $\Delta=1.5$
and $D=6.60J$; (f) $\Delta=-0.5$ and $D=-1.50J$.
}
\end{figure*}

\appendix

\section{\label{sec:appendix A}Model with single-ion anisotropy}

To further explore the physics of bicriticality in a local-moment
model, it is instructive to consider, in addition to the exchange
anisotropy, a single-ion anisotropy of the form $-D\sum_{i}(S_{i}^{z})^{2}$.
While such a term is absent from a spin-1/2 model, it is generically
present for spins $S\geq1$ in a tetragonal environment.

For illustration, we consider a model with ferromagnetic nearest-neighbor
exchange only, with the Hamiltonian: \begin{equation}
\mathcal{H}=-\sum_{\left\langle i,j\right\rangle }\sum_{\alpha}J^{\alpha}S_{i}^{\alpha}S_{j}^{\alpha}-D\sum_{i}\left(S_{i}^{z}\right)^{2}\label{h2}\end{equation}
 where $D>0$ $\left(D<0\right)$ favors spin alignment along the
c axis (in the ab plane). For the exchange anisotropy, we stick to
the parametrization $J^{x}=J^{y}=J\left(1+\Delta\right)$ and $J^{z}\equiv J$.
Studies of the model \eqref{h2} have appeared before in the literature,\cite{freire04,selke13}
and we complement these results here.

%%%%%%%%%%%%%%%%%%%%%%%%%%%%%%%%%%%%%%%%%%%%%%%%%%%%%%%%%%%%%%%%%%%%%%%

\subsection{Phase diagram}

The competition between the single-ion anisotropy $D$ and the exchange
anisotropy $\Delta$ in Eq.~\eqref{h2} leads to phase diagrams as
shown in Fig.~\ref{fig:Phase-diagram} (Fig.~\ref{fig:Phase-diagram}b
matches within error bars the corresponding result in Ref.~\onlinecite{freire04}).
At $T=0$ there is a first-order phase transition between in-plane
and c-axis order at $D=zJ\Delta/2$, where $z$ is the lattice coordination
number. This transition continues to finite $T$ and terminates at
at bicritical point.\cite{selke13,shapira70,selke11} To efficiently sample all spin
configurations close to this first-order phase transition, we employ
the parallel-tempering algorithm.\cite{partemp_jpsj,fiore08} The thermal
transition at the bicritical point is in the 3d Heisenberg universality
class, i.e., the O(3) symmetry is restored here. We notice that when
crossing the first-order line upon heating up the system, we always
favor the state which is the ground state for $D=0$, which has a
higher entropy because both spin angles are unlocked.

\subsection{Susceptibilities}

We now turn to the behavior of the anisotropic magnetic susceptibilities,
Fig. \ref{fig:chi}. Motivated by the experimental results to \yrsc,
we only consider magnetic order along the c axis. For all results
in Fig. \ref{fig:chi}, we consider $g_{ab}=g_{c}$ and the effects
of the single ion anisotropy are encoded by the parameter $D$. In
Figs. \ref{fig:chi}a-c we see that as we approach the bicritical
point, the fluctuations in $\chi_{ab}$ are enhanced. Interestingly,
in \ref{fig:chi}a-b, we even have $\chi_{ab}>\chi_{c}$ at intermediary
temperatures above $\TC$, with the two curves crossing only very
close to the transition. This difference between $\chi_{ab}$ and
$\chi_{c}$ is further enhanced if we increase the single-ion anisotropy
$D$, Figs.~\ref{fig:chi}d and \ref{fig:chi}e (in both of them
we keep the ratio $D/\Delta$ the same as in Fig.~\ref{fig:chi}a
so we do not move too far away from the bicritical point). Finally
in Fig. \ref{fig:chi}f we consider both $D$ and $\Delta$ negative
and we always have $\chi_{c}>\chi_{ab}$ for $T>\TC$.

Therefore, we see that model \eqref{h2} naturally describes the crossing
of the susceptibilities above $\TC$ as a result of the proximity
to a bicritical point. While it is true that the splitting between
$\chi_{ab}$ and $\chi_{c}$ above the crossing point is small, it
is still remarkable that a purely local moment description is able
to capture such crossover, an observation which motivates us to associate
this crossing to the proximity of a bicritical point.

%%%%%%%%%%%%%%%%%%%%%%%%%%%%%%%%%%%%%%%%%%%%%%%%%%%%%%%%%%%%%%%%%%%%%%%

\end{document}